# Surface termination control of charge transfer and band alignment across a semiconductor-crystalline oxide heterojunction

M. Chrysler[1], J. Gabel[2], T.-L. Lee[2], Z. Zhu[3], T. C. Kaspar[4], P. V. Sushko[4], S. A. Chambers[4], and J. H. Ngai[1]

[1]*Department of Physics, University of Texas-Arlington, Arlington, TX 76019, USA*
[2] *Diamond Light Source, Ltd., Harwell Science and Innovation Campus, Didcot, Oxfordshire OX11 0DE, England, United Kingdom*
[3]*Environmental Molecular Sciences Laboratory Pacific Northwest National Laboratory, Richland, WA 99352, USA*
[4]*Physical Sciences Division, Physical & Computational Sciences Directorate, Pacific Northwest National Laboratory, Richland, WA 99352, USA*

Charge redistribution across heterojunctions has long been utilized to induce functional response in materials systems. Here we examine how the composition of the terminating surface affects charge transfer across a heterojunction consisting of Si and the crystalline complex oxide $SrTiO_3$. Itinerant electrons in Si migrate across the interface toward the surface of $SrTiO_3$ due to surface depletion. The electron transfer in turn creates an electric field across the interface that modifies the interfacial dipole associated with bonding between $SrTiO_3$ and Si, leading to a change in the band alignment from type-II to type-III. By capping the $SrTiO_3$ surface with ultra-thin ($\leq$ 1 nm) layers of BaO, SrO or $TiO_2$, charge transfer across the interface can be weakened or inhibited. Ab initio modeling implicates the adsorption of oxygen as driving surface depletion in $SrTiO_3$. The electronic coupling between the surface and buried interface expands the functionality of semiconductor-crystalline oxide heterojunctions.

I.  INTRODUCTION

Semiconducting heterojunctions exhibiting built-in electric fields due to charge transfer underpin the functionality of virtually all electronic device technologies. While such heterojunctions have traditionally been comprised of covalently bonded semiconductors, recent interest has turned to hybrid heterojunctions in which one side of the heterojunction is an ionically bonded crystalline oxide [1,2]. Such hybrid heterojunctions potentially enable the ideal characteristics of both covalently bonded and ionically bonded semiconductors to be combined. Heterojunctions possessing complementary covalent and ionic characteristics are of potential interest for a variety of applications, including those in which the surface of the oxide is exposed to reactive terrestrial conditions [3,4]. Understanding how the electronic behavior of hybrid heterojunctions evolves under such conditions is necessary. However, to date, even an understanding of how the electronic properties of hybrid heterojunctions evolve with mere exposure to ambient conditions is lacking.

In this regard, surfaces play a pivotal role in determining the electronic behavior of semiconducting thin films and heterojunctions. Surfaces break the translational symmetry of the crystal lattice, which leads to the formation of undercoordinated atoms that can attract adsorbates, or form states that trap charge, or give rise to structural reconstructions at the surface [5]. Itinerant carriers in the bulk, if present, can contribute to this interplay between dangling surface bonds, adsorbates, and surface reconstructions. In conventional semiconducting materials, the phenomenon of surface depletion arises as itinerant carriers are drawn to the surface away from their dopant ions in the bulk, thereby forming space charge and band bending near the surface [6].

Here we explore the interplay between surface depletion and charge transfer across the archetype semiconductor-crystalline oxide heterojunction comprised of Si and $SrTiO_3$ (STO). We find that itinerant electrons in the former migrate across the interface toward the surface of the latter due to surface depletion. This surface depletion has a dramatic effect on the electronic structure of the interface, as electrons transferred from Si to STO create space charge that modifies the interfacial dipole associated with bonding between the two materials, leading to a change in the band alignment from type-II, in which the conduction band minimum (CBM) of STO is above the valence band maximum (VBM) of Si, to type-III, in which the STO CBM is below the Si VBM. We find that surface depletion is dependent on the composition of the terminating layer at the oxide surface. Ultra-thin capping layers of alkaline earth oxide BaO inhibit charge transfer across the interface, as manifested in electrical transport and hard x-ray photoelectron spectroscopy (HAXPES) measurements. We find that even a sub-monolayer (ML) of additional SrO or $TiO_2$ deposited on the as-grown STO surface can significantly alter electrical behavior. Density functional theory (DFT) modeling reveals that dissociated oxygen adsorbed on the surface of the STO upon exposure to ambient conditions likely drives surface depletion. Coupling of the electronic degrees of freedom of the surface with the electronic behavior of the buried interface opens additional pathways to exploit semiconductor-crystalline oxide heterojunctions in applications.

## II. EXPERIMENTAL DETAILS

The epitaxial thin films of STO were grown through co-deposition of SrO and $TiO_2$ fluxes on nominally undoped, Czochralski-grown (CZ) Si substrates (Virginia Semiconductor) using oxide molecular beam epitaxy (MBE), under conditions that are described in detail elsewhere [2,7]. The Si substrates have dimensions of 1.4 × 1.4" (2" diagonal) and were diced from larger 4" diameter wafers. Here we present results from epitaxial STO/Si heterojunctions for which the surface of the STO was either left as-grown, or capped *in situ* immediately after film growth with: (i) 1 nm of epitaxial BaO, (ii) 0.7 MLs of epitaxial SrO, and (iii) 0.7 MLs of $TiO_2$. The epitaxial BaO, SrO, and $TiO_2$ layers were deposited at a substrate temperature of ~ 400 °C in a chamber background pressure of ~ $2 \times 10^{-7}$ Torr $O_2$. We chose these capping layers to elucidate the effects of surface composition on the electronic behavior of the buried interface. Specifically, the 1 nm thick BaO capping layer is a wide band gap insulator that would screen the STO substate from the effects of surface adsorbates while 0.7 ML of SrO and 0.7 ML of $TiO_2$ change the fractions of the STO surface termination with SrO and $TiO_2$ planes.

Electrical transport measurements of the heterojunctions were carried out in the Van der Pauw geometry in a Quantum Design Dynacool system as a function of temperature and magnetic field. Sheet resistance and Hall measurements were performed using a Keithley 2400 Sourcemeter and Keithley 2700 Multimeter fitted with a Keithley 7709 Matrix Module to enable multiplexing between all the Van der Pauw lead configurations. Electrical contacts to the samples were made using Al wedge bonding (Westbond) to the corners of diced 5×5 $mm^2$ samples. Two-point current-voltage measurements confirm Ohmic behavior in the contacts. The Hall data were fitted to two- or three-carrier models that account for electron conductivity in the STO in conjunction with electron and hole conductivity in the Si. The fits to the Hall data were constrained such that the resulting carrier densities and mobilities would yield the corresponding zero-field sheet resistance obtained at the same temperature as the Hall resistance. We additionally constrain the mobilities to be close to those reported for bulk Si and STO films.

HAXPES measurements were made at the Diamond Light Source (UK) on Beamline I09. An x-ray energy of 5.9 keV was selected using a Si(111) double crystal monochromator followed by a Si(004) channel-cut high resolution monochromator. The films were found to be highly resistive. To avoid surface charging, we reduced the incident x-ray flux until further reduction did not yield a measurable shift over time in the binding energy of the Ti $2p_{3/2}$ core level. A Scienta Omicron EW4000 high-energy hemispherical analyzer had to be set to a 500 eV pass energy to compensate for the low x-ray flux, resulting in an overall experimental resolution of ~ 500 meV, as judged by fitting a Fermi function to the Fermi edge of a gold foil. The binding energy scale was calibrated using the Au $4f_{7/2}$ core-level along with the Fermi edge of a gold foil. All spectra were measured with an angle of x-ray incidence of approximately 10° relative to the surface plane. The angle between the analyzer lens axis, which lays in a horizontal plane, and the incoming x-rays was 93°. Thus, the photoelectron take-off angle was 13° off normal. Soft x-ray XPS was also performed at a pass energy of 200 eV using x-ray energies that are ~470 eV greater than the core-level binding energies of interest in order to maintain a constant probe depth of ~ 2 nm [8]. The total energy resolution in the soft x-ray measurements was ~300 meV.

In order to extract band-edge profiles, we utilize an algorithm we designed to deconvolve heterojunction core-level line shapes into a set of layer-resolved spectra that yields a binding energy profile for each component material in the heterostructure [9].

For the DFT modeling of O bonding to the n-type STO (n-STO) and BaO surfaces, the BaO film grown on the n-STO (001) surface was represented using the periodic slab model. The STO part of the slab was 7 unit cells thick, while the thickness of the BaO film was varied from one to four atomic planes. The lateral supercell of the slab corresponds to the 2×2 crystallographic perovskite cell; the lateral supercell parameters were fixed at a=b=7.81 Å, which corresponds to the bulk STO lattice constant of 3.905 Å. The out-of-plane supercell parameter was set to 50 Å, which leaves the vacuum gap of at least 14 Å depending on the BaO thickness. The binding energy ($E_b$) of the oxygen atom at the n-STO and BaO/n-STO surfaces was calculated with respect to the half of the gas-phase $O_2$ molecule energy as $E_b$ = E(Slab) + 1/2E($O_2$) – E(Slab + O); i.e., $E_b$ is positive if O binding to the surface is energetically preferred. Atoms in the SrO plane furthest from the BaO/n-STO interface were fixed at the sites corresponding to the ideal bulk lattice. The total energy of each system was minimized with respect to the internal coordinates of all other atoms. The calculations were performed using the VASP package [10,11] and the PBEsol exchange-correlation functional [12]. Projector-augmented wave potentials were used to approximate the effect of the core electrons [13]. A Γ-centered 2×2×1 $k$-mesh was used for Brillouin-zone integration in the structure optimization calculations; a 12×12×1 $k$-mesh was used for calculations of the density of states (DOS). The plane-wave basis-set cutoff was set to 500 eV. The total energy convergence criterion was set to $10^{-5}$ eV. The charge population analysis was performed using the method developed by Bader [14,15].

Time-of-flight secondary ion mass spectroscopy (ToF-SIMS) measurements were performed with a TOF.SIMS5 instrument (IONTOF GmbH, Münster, Germany) using a dual beam depth profiling strategy. A 1.0 keV $Cs^+$ beam (~45 nA) was used for sputtering. The $Cs^+$ beam was scanned over a 300 × 300 $\mu m^2$ area. A 25.0 keV $Bi_3^+$ beam (~ 0.57 pA) was used as the analysis beam to collect SIMS depth profiling data. The $Bi_3^+$ beam was focused to be ~5 microns in diameter and scanned over a 100 × 100 $\mu m^2$ area at the center of the $Cs^+$ crater.

### III. RESULTS & DISCUSSION

The STO/Si heterojunction capped with 1 nm of BaO exhibits strikingly different electrical transport behavior and electronic structure than the as-grown heterojunction. The as-grown STO/Si heterojunction exhibits lower sheet resistance $R_s$ than the BaO-capped heterojunction, along with an unusual non-monotonic anomaly (arrow), as shown in Fig. 1(a). Hall measurements reveal non-linear behavior and a crossover in sign in $R_{xy}$ that occurs at a temperature near the anomaly in $R_s$ (Fig. 1(c)), indicating the emergence of a high-mobility hole-gas in Si which conducts in parallel with the *n*-type carriers in the STO. Fits to the $R_{xy}$ data using 2- and 3-carrier models yield the sheet densities $n_{e,STO}$ and mobilities $\mu_{e,STO}$ of the electrons in the STO (Fig. 1(f)), as well as the sheet densities $n_{e/h,Si}$ and mobilities $\mu_{e/h,Si}$ of both holes (hollow) and electrons (solid) in the Si (Fig. 1(g)). Indeed, an electron and a hole channel are needed in the Si to properly fit the non-linearity in the $R_{xy}$ data at 400 K. The additional electrons in the Si at 400 K are attributed to intrinsic carriers that are thermally excited. In contrast, the STO/Si heterojunction capped with 1 nm of epitaxial BaO exhibits a higher $R_s$ and insulating behavior ($dR_s/dT < 0$) at all temperatures (Fig. 1(a)). While $R_{xy}$ (Fig. 1(b)) also exhibits non-linear behavior, analysis indicates that in the BaO capped sample only *n*-type carriers are present, comprised of electrons in the STO that exhibit low mobilities (< 10 cm$^2$/Vs, Fig. 1(d)), and electrons in the Si that exhibit much higher mobilities (> 1000 cm$^2$/Vs, Fig. 1(e)).

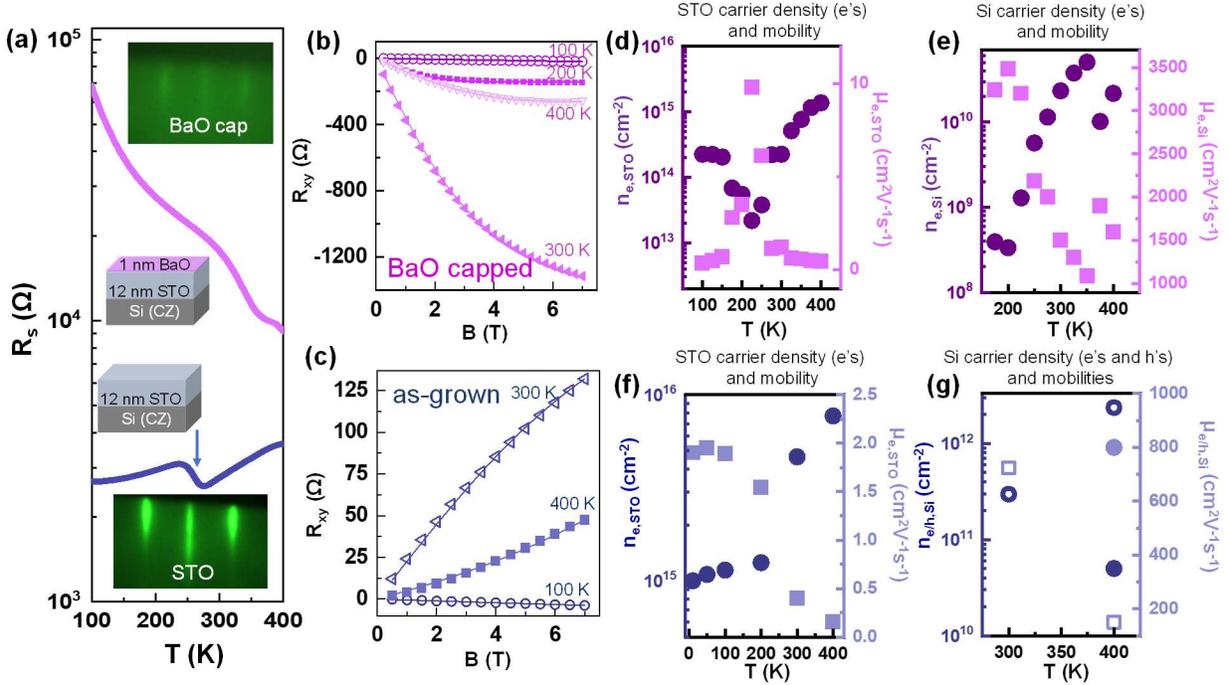

Fig. 1. (a) $R_s$ for the 1 nm BaO capped and the as-grown 12 nm STO/CZ-Si heterojunctions. Insets show RHEED of the as-grown and 1 nm BaO capped heterojunctions. (b) and (c) show $R_{xy}$ for the 1 nm BaO capped and as-grown heterojunctions, respectively. Raw data is shown as symbols and fits to a 2- or 3-carrier model are shown as lines. Note the crossover in the sign of $R_{xy}$ for the as-grown heterojunction indicating the formation of a hole gas. (d) and (e) show carrier densities and mobilities in the STO and Si, respectively, for the 1 nm BaO capped heterojunction extracted from fits to $R_{xy}$. (f) and (g) show carrier densities and mobilities in the STO and Si, respectively, for the as-grown heterojunction extracted from fits to $R_{xy}$. Closed shapes indicate electron carrier densities and mobilities, while open shapes indicate hole carrier densities and mobilities.

HAXPES measurements and related modeling reveal the electronic structures that give rise to the electrical transport behavior of the as-grown and BaO capped heterojunctions. The Ti 2p spectrum of the as-grown heterojunction exhibits prominent lower valence features, e.g., Ti$^{3+}$, (Fig. 2(a)), which corroborates the large $n_{e,STO}$ deduced by transport. Furthermore, the Ti 2p and Sr 3d core-level spectra exhibit asymmetries toward higher binding energy, while the Si 2p spectrum exhibits asymmetry toward lower binding energy for the as-grown heterojunction (arrows, Fig. 2a). Such asymmetries are consistent with the presence of built-in electric fields. As described in Section II, these spectra can be fitted (Fig. 2(b) and 2(c)) to extract the valence band edge profile across the heterojunction (Fig. 2(d)). Such fitting reveals a type-III band alignment in the as-grown heterojunction, in which the valence band of Si is above the conduction band of STO, thereby enabling a hole-gas to emerge, as found in Hall measurements. In contrast, the lower valence features in the Ti 2p spectrum are much weaker for the BaO capped heterojunction (Fig. 2(a)), and all core-level spectra are largely symmetric. Fits to the spectra (Fig. 2(c)) indicate that the BaO capped heterojunction exhibits a type-II band alignment, in which the conduction band of STO is situated below (above) the conduction (valence) band of Si (Fig. 2(d)) [16].

As described in detail elsewhere [2,7], the high $n_{e,STO}$, hole-gas and the built-in electric fields in the as-grown STO/Si heterojunction arise from the transfer of itinerant electrons from Si to STO. The itinerant electrons originate from O impurities in the Si that act as *n*-type donors. The O impurities stem from two sources, namely, O that has diffused into Si during the epitaxial growth of STO, and O that is intrinsically present in the CZ-grown Si substrate [17]. The transfer of itinerant electrons to STO creates space charge and in turn a built-in electric field across the interface. In a conventional semiconducting heterojunction, the transfer of itinerant charge and the formation of space charge generally leaves the type of band alignment across the interface unchanged. However, for our STO/Si heterojunctions, the electric field across the heterojunction that arises from the transfer of electrons alters the interfacial dipole associated with epitaxial bonding between STO and Si [18]. The change in interface dipole moment in turn causes the offset between the Si and STO valence bands to increase, thereby inducing the band alignment to change from type-II to type-III [7]. The hole-gas in the Si is a manifestation of inversion, as the O impurity donors are depleted under a type-III alignment, and the valence band near the interface is pulled upwards towards the Fermi Energy $E_F$. In contrast, the BaO-capped heterojunction exhibits much weaker charge transfer and a type-II band alignment. Since both capped and as-grown heterojunctions were grown under identical conditions and on identical Si substrates and exhibited virtually identical O impurity depth profiles, as determined by ToF-SIMS (Fig. S1) [19], the difference in their electronic behavior cannot be attributed to a difference in density of O donors in the Si substrate [7].

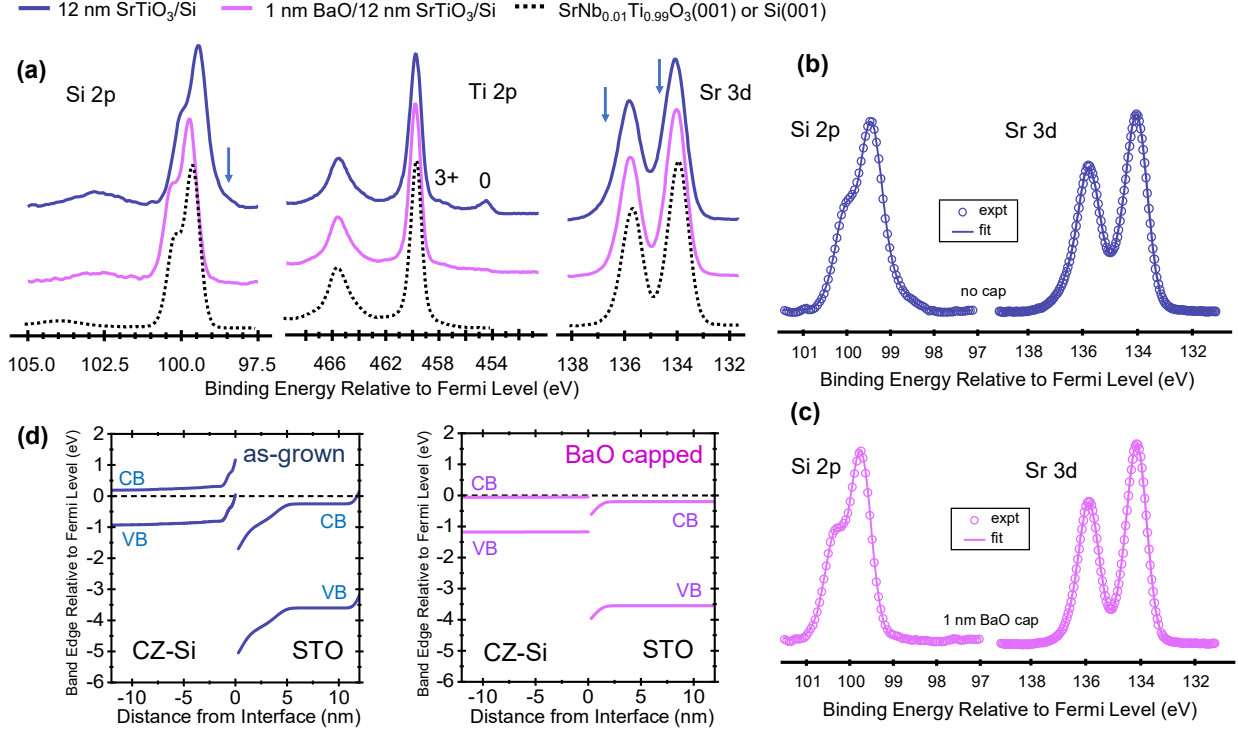

Fig. 2. (a) Core-level Si 2p, Ti 2p and Sr 3d spectra for as-grown (blue) and 1 nm BaO capped (magenta) 12 nm STO/CZ-Si heterojunctions. Also shown are reference spectra (black) from single-crystal Si(001) and $SrNb_{0.01}Ti_{0.99}O_3(001)$ substrates. (b) and (c) show fits to the Si 2p and Sr 3d spectra for the as-grown and 1 nm BaO capped heterojunctions, respectively. (d) Band edge profiles across the heterojunctions extracted from the fits in (b) and (c). Note the type-III (type-II) alignment in the as-grown (1 nm BaO capped) heterojunction.

As the presence of a BaO capping layer is the only principal difference between the two heterojunctions, we consider the effects of the surface on the electrical behavior and electronic structure of the heterojunctions. We note that the upward band-bending at the surface of the as-grown STO/Si heterojunction revealed by HAXPES (Fig. 2(d)) is reminiscent of surface depletion, which causes carriers in the bulk to be pulled to the surface, thereby creating space charge and a built-in electric field. In contrast, the BaO-capped heterojunction does not exhibit any measurable band-bending at the surface based on HAXPES.

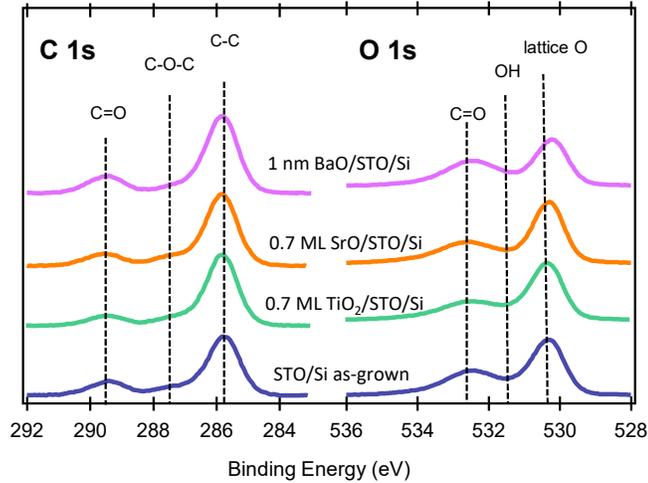

Fig. 3. Soft x-ray C 1s and O 1s core-level spectra for the 1 nm BaO, 0.7 ML SrO, 0.7 ML TiO$_2$ and as-grown 12 nm STO/CZ-Si heterojunctions. Contributions from various surface adsorbates are indicated with dashed lines.

The adsorption of various species on an oxide surface upon exposure to ambient conditions is a key mechanism that can drive surface depletion [5]. On the premise that the difference in behavior between the as-grown and BaO-capped heterojunctions can be attributed to a difference in adsorbates, more surface-sensitive soft x-ray XPS measurements were carried out. Figure 3 shows the C 1s and O 1s spectra in which contributions from the various adsorbed species are indicated. The as-grown and BaO capped heterojunctions do not exhibit any significant difference in the kinds of functional groups that are adsorbed on their surfaces, as indicated by the similar binding energies and intensities for the various spectral features. Furthermore, adsorbed oxygen, which is a key adsorbate that will be discussed below, cannot be resolved using XPS as the O 1s spectroscopic signature of dissociated $O^{2-}$ strongly overlaps with that of oxygen in the lattice. Thus, XPS alone does not yield clear insight.

Therefore, we turn to ab initio calculations to gain further insight into the changes in electronic structure resulting from surface adsorption. Here we focus on the adsorption of oxygen, which is a strong and ubiquitous electron scavenger that has been established to induce surface depletion in STO by reacting with itinerant electrons to form $O^{2-}$ [5,20-26]. For our heterojunctions, such itinerant electrons stem not only from residual oxygen vacancies $V_O$ in the STO, but also from oxygen impurities in the Si that act as *n*-type donors.

We first compare O adsorption on the $TiO_2$-terminated STO surface with that on the BaO surface in the BaO-capped STO. Figure 4(a) shows a schematic of the slab model used to investigate oxygen adsorption. To mimic the effect of itinerant electrons in the n-STO arising from electron transfer from Si or residual O vacancies, we introduce a single $V_O$ located in an SrO plane. Two locations of the $V_O$ were considered: the first SrO plane from the BaO/STO interface (denoted as $V_O$-S1) and the SrO plane ~2 nm away from the interface (denoted as $V_O$-S5).

We consider several sites at which oxygen can potentially bind to the $TiO_2$-terminated STO surface or the BaO surface. For the case of the as-grown STO surface that is terminated with a $TiO_2$ layer, oxygen can bind directly above a surface $Ti^{4+}$ ion (this configuration is denoted by OT in Fig. 4(b) and 4(c)), or directly above a surface $O^{2-}$ ion to form a peroxy-like species $O_2^{2-}$ (OO-STO in Fig. 4(b) and 4(c)). For the case of the BaO capped heterojunctions, oxygen can similarly bind to a surface $O^{2-}$ ion (denoted by OO-BaO in Fig. 4(b) and 4(c)), or occupy a site located between two $Ba^{2+}$ ions (denoted by OBB in Fig. 4(b) and 4(c)). The configuration of the adsorbed oxygen directly above a $Ba^{2+}$ ion was also examined but found to be unstable for all BaO thicknesses; thus, this configuration is not discussed further.

The calculated oxygen binding energies for these sites are shown in Fig. 4(b). For the as-grown, $TiO_2$-terminated STO surface (corresponding to N = 0 BaO layers), oxygen binds most strongly to Ti ($E_b \approx 3$ eV for OT) for the scenario in which the oxygen vacancy is located in the SrO layer immediately below the terminating $TiO_2$ surface, and directly below the Ti site at which oxygen is adsorbed ($V_O$-S1). As the distance between the $V_O$ and the terminating $TiO_2$ surface increases ($V_O$-S5), the oxygen binding energy decreases to < 1 eV. In contrast, oxygen situated above a surface $O^{2-}$ (OO-STO) is unstable, as the binding energy is negative. While the OO-STO configuration is unfavorable, OO-BaO becomes energetically stable as BaO layers are added. The binding energy increases from $E_b \sim 0$ eV for a single plane of BaO, to 1 eV for two BaO planes, and saturates at that energy for thicker BaO films. In contrast, the binding energy for oxygen adsorbed at the OBB-BaO site weakens with increasing BaO thickness (e.g., $E_b \sim 0$ eV for the 4-layer thick BaO film).

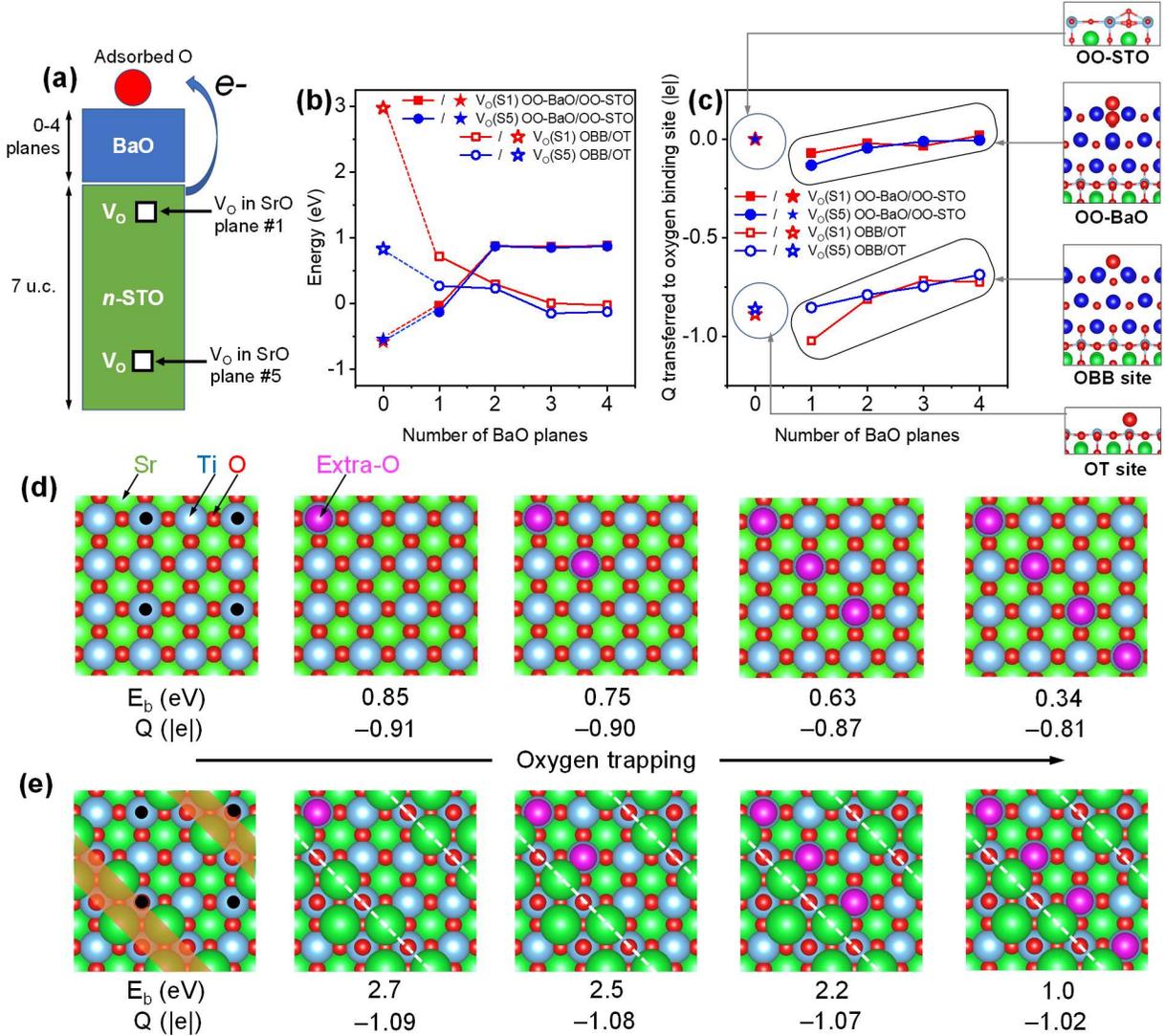

Fig. 4. Ab initio modelling of oxygen adsorption. (a) Schematic of the slab used in our calculations and the locations of oxygen vacancies used to mimic n-STO. (b) Binding energies for oxygen at various sites on the $TiO_2$ and BaO surfaces, as a function of the number of BaO layers N, where N = 0 corresponds to the $TiO_2$ surface. (c) Charge transferred from STO to O binding site as a measure of electron scavenging, along with schematics of the various adsorption sites. Note that O adsorbed at the energetically stable site of OO-BaO does not scavenge charge from STO. (d) Binding energies ($E_b$) and the amount of trapped charged (Q) for a chain of oxygen atoms of varying length on the $TiO_2$ surface. (e) $E_b$ and Q of the same chain of oxygen atoms in the presence of surface excess SrO modeled here as SrO strips (shaded regions in (e)), which introduce steps on the surface that result in the adsorbed oxygen to be coordinated by $Sr^{2+}$ and $Ti^{4+}$. Black dots in (d,e) indicate the lateral coordinated of $V_O$ located immediately below the topmost $TiO_2$ plane.

To quantitatively correlate the binding energies with the amount of electron charge transferred from n-STO to the adsorbed oxygen, we define the transferred charge as the difference between the total number of electrons attributed to the BaO film and BaO+O system, i.e., before and after O adsorption, respectively. The values of these charges for all configurations considered are plotted in Fig. 4(c). We note that the charges calculated for the as-grown $TiO_2$-terminated n-STO and BaO capped n-STO systems can be compared only qualitatively because of the differences in the definition of the atomic volume used by the Bader analysis for these two systems. Analysis of these trends suggests that oxygen binding at the cation sites (OT and OBB) is accompanied by the transfer of 0.7 – 1 electrons, which is consistent with scavenging behavior. The amount of charge transferred to the oxygen adsorbate shows little dependence on the location of the $V_O$ in STO (Fig. 4(c)). However, only OT is energetically stable while OBB becomes less stable with increasing BaO thickness. Examination of the projected density-of-states (DOS) provides further insight. Whereas binding at the OT site traps itinerant electrons at the surface in a deep gap state (Fig. S2(a)), binding at the OBB site instead produces a shallow gap state (Fig. S2(b)). For comparison, oxygen binding at either the OO-BaO site or OO-STO site does not induce the transfer of charge from STO (Fig. S2(c) and S2(d)), but is instead reminiscent of the formation of peroxy $O_2^{2-}$ species in the case of interstitial oxygen atoms.

Thus, our ab initio calculations for the $TiO_2$-terminated and BaO-capped n-STO surfaces indicate that capping STO with BaO inhibits surface depletion. The binding of oxygen to the energetically stable OO-BaO sites does not draw itinerant electrons from n-STO. We also note that electron transport from the n-STO to the surface is kinetically inhibited due to the sizeable offset between the conduction bands of STO and BaO, in which the latter is situated above the former in energy (Fig. S2(e)). In contrast, for the $TiO_2$ terminated STO surface, the binding of oxygen to the energetically stable OT sites draws electrons to the surface. We argue that this scavenging of electrons to the STO surface induces a self-reinforcing transfer of electrons across the interface. As electrons are pulled from Si to the STO surface, the resulting space charge across the interface modifies the interfacial dipole and increases the valence band offset (VBO). The increasing VBO in turn promotes the transfer of more electrons from Si to STO.

Up to this point, we considered idealized $TiO_2$-termination of the as-grown n-STO. However, the surface of our as-grown heterojunctions likely exhibits a mixture of SrO and $TiO_2$ terminations. Mixed termination could arise from drift in the stability of Sr and Ti fluxes over the duration of the growth and due to dislocations that thread to the surface. Such dislocations stem from anti-phase boundaries that emerge at steps on the Si(100) surface and these are incommensurate in height with the lattice constant of STO [27].

Accordingly, we have also performed ab initio calculations for a SrO-terminated surface and a $TiO_2$-terminated surface that is partially covered by quasi-one-dimensional SrO islands, in which the island step edge is parallel to the [100] direction of STO. For the SrO-terminated surface, we find that the binding energies are comparable to those of having 1 ML coverage of BaO on top (Fig. 4(b)): ~0.1 eV for the OSS site, which is structurally equivalent to the OBB site on the BaO-capped STO and ~0.2 eV for the OO-STO site. To quantify the effect of the surfaces with mixed termination on the $E_b$, we first isolate the effect of SrO steps. To this end, we compare the O binding energies calculated for the same amount and the same locations of the adsorbed O (shown with magenta in Fig. 4(d) and 4(e)) with and without SrO steps. Figure 4(d) shows a view of the top surface of our slab, in which we have introduced 4 $V_O$'s (amounting to 0.25 per crystallographic lateral cell) situated in the SrO layer immediately below the surface layer of $TiO_2$; lateral positions of the $V_O$'s are shown with black dots in Fig. 4(d) and 4(e). The binding energies of each additional O species and corresponding charge that is transferred from the n-STO are indicated in Fig. 4(d). Note that the binding energies for the oxygen are smaller than for the case of binding at an OT site discussed above (Fig. 4(b)), because the $V_O$'s are laterally displaced from the adsorbed oxygen, and not immediately below them. As diagonal strips of SrO covering 50% of the surface are added on top (brown trapezoids), the binding energies for adsorbed O increases, as indicated in Fig. 4(e). We find that this enhancement in binding energy is not unique to the SrO steps but is, in fact, germane to a variety of SrO structures, such as single SrO molecules, square $Sr_2O_2$ islands, and zig-zag SrO chains (see Fig. S3 in the Supplementary Information for more details). This enhancement in binding energy on surfaces with SrO structures on top can be generally attributed to an increase in the coordination of the adsorbed oxygen, as bonding occurs not just to a $Ti^{4+}$ ion below but to an adjacent $Sr^{2+}$ ion as well.

To experimentally explore the effect that mixed termination has on electron transfer across the interface, we compare the results of the as-grown STO/Si heterojunction with STO/Si heterojunctions that have an additional 0.7 sub-ML of SrO or $TiO_2$ deposited on top. Like the as-grown and BaO capped heterojunctions, the SrO and $TiO_2$ sub-ML capped heterojunctions exhibit virtually identical surface adsorbates (Fig. 3). Figure 5(a) shows $R_s$ for the SrO sub-ML capped heterojunction, showing higher $R_s$ for all temperatures in comparison to the as-grown STO/Si heterojunction (Fig. 1(a)), as well as the TiO$_2$ sub-ML capped heterojunction. No crossover in sign is observed in $R_{xy}$ for the SrO sub-ML capped heterojunction, indicating the absence of a robust hole-gas. Despite the persistent negative sign of $R_{xy}$, a hole-gas does emerge albeit at 400 K, as the addition of a hole channel was required to properly fit the $R_{xy}$ data at that temperature (Fig. 5(b) and Fig. S4). Nonetheless it is clear that sub-ML deposition of SrO significantly weakens electron transfer from Si to STO, as also evident in comparing $n_{e,STO}$ between the sub-ML deposited (Fig. S4(a)) SrO heterojunction, to the as-grown heterojunction (Fig. 1(f)). Fits to the Si 2p and Sr 3d HAXPES spectra (Fig. S5) corroborate the transport measurements, as a type-II band alignment is revealed (Fig. 5(d)).

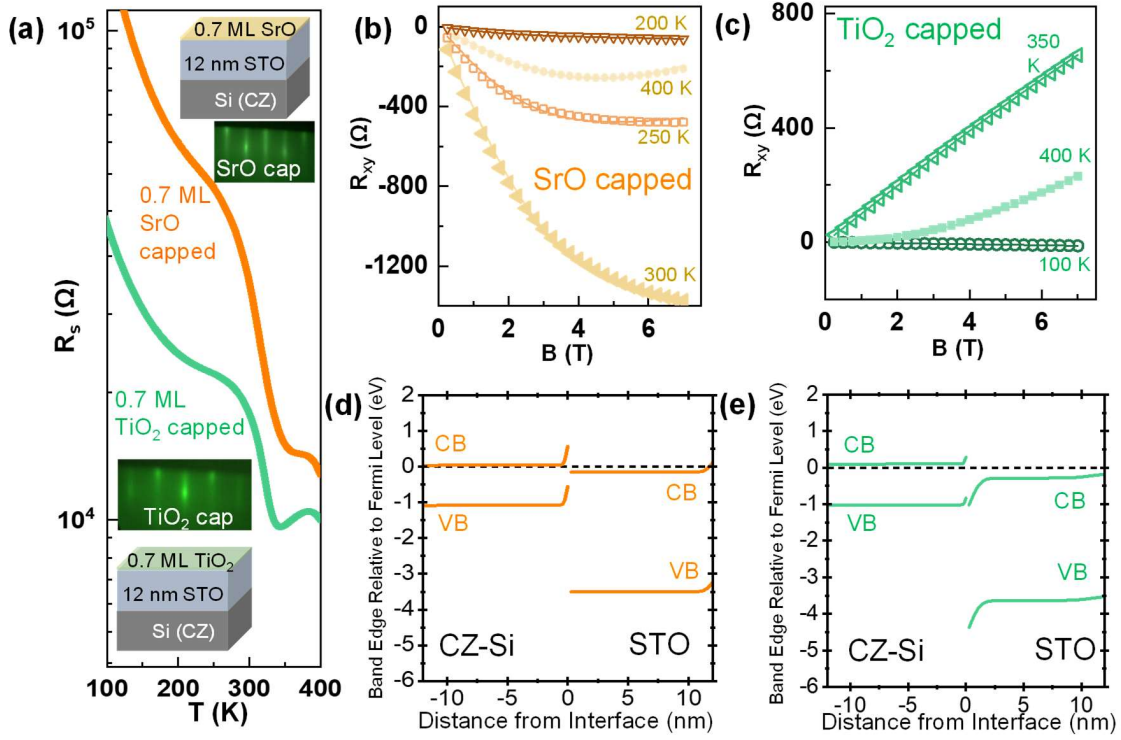

Fig. 5. (a) $R_s$ for the 0.7 ML SrO and 0.7 ML TiO$_2$ capped 12 nm STO/CZ-Si heterojunctions. Insets show RHEED for both heterojunctions. (b) and (c) show $R_{xy}$ for the 0.7 ML SrO and 0.7 ML TiO$_2$ capped 12 nm STO/CZ-Si heterojunctions, respectively. (d) and (e) show band edge profiles for the 0.7 ML SrO and 0.7 ML TiO$_2$ capped STO/CZ-Si heterojunctions, respectively.

In contrast, the sub-ML TiO$_2$ capped heterojunction exhibits lower $R_s$ and a crossover in sign of $R_{xy}$ from negative to positive, indicating a higher $n_{e,STO}$ than the sub-ML SrO capped heterojunction and the presence of a robust hole-gas, respectively. However, we note that the sub-ML TiO$_2$ capped heterojunction exhibits a higher $R_s$ as well as lower $n_{e,STO}$, in comparison to the as-grown heterojunction. On the premise that larger O binding energies enhance surface depletion by promoting electron localization on the adsorbed oxygens, it follows from our ab initio calculations that the as-grown heterojunction likely has the largest areal density of exposed step edges. Capping with TiO$_2$ reduces the areal density of these exposed edges and, therefore, the amount of excess oxygen bonded through double coordination. While capping with SrO also reduces the areal density of step edges, SrO further hinders surface depletion as binding at the energetically stable OO site does not draw electrons from the bulk, similar to capping with BaO.

We remark on the relationship between the electric field at the STO surface due to surface depletion and the electric field at the interface, which forms a well that confines the electrons transferred from Si to STO. The electric field at the interface does not appear to be a continuation of the field due to surface depletion, as our fits to the HAXPES spectra of the uncapped heterojunction indicate a region with in which the bands are largely flat between the surface and the interface (Fig. 1(d)). The transferred electrons accumulate near the interface due to the potential of the ionized donors in the Si, leading to the observed band bending.

Finally, we remark on the band gap of STO shown in the band-diagrams derived from HAXPES. Spectroscopic ellipsometry (Fig. S6) indicates the bulk indirect band gap is 3.35 eV for a typical 12 nm thick STO film on grown Si. The enhancement of the band gap of STO grown on Si relative to the bulk value (3.25 eV) is likely attributable to strain and residual disorder, such as dislocations [28,29].

## IV. CONCLUSIONS

In summary, we demonstrate that the surface of an oxide can be electrically coupled to the electronic structure of the buried interface in semiconductor-crystalline oxide heterojunctions. Surface depletion in STO/Si heterojunctions drives not only charge transfer across the interface, but also a change in band alignment when coupled with the interfacial dipole. Ab initio calculations implicate the adsorption of oxygen as a likely mechanism responsible for the electrical behavior observed in our heterojunctions.

ACKNOWLEDGEMENTS

This work was supported by the National Science Foundation (NSF) (Awards Nos. DMR-1508530, CMMI-2132105). HAXPES analysis and ab initio modeling work was supported by the U.S. Department of Energy, Office of Science, Basic Energy Sciences, Materials Sciences and Engineering Division, Synthesis and Processing Science program, FWP 10122. This research used resources of the National Energy Research Scientific Computing Center, a DOE Office of Science User Facility supported by the Office of Science of the U.S. Department of Energy under Contract No. DE-AC02-05CH11231 using NERSC award BES-ERCAP0021800. A portion of this research was performed on a project (Award DOI: 10.46936/cpcy.proj.2021.60271/60008423) from the Environmental Molecular Sciences Laboratory, a DOE Office of Science User Facility sponsored by the Biological and Environmental Research program under Contract No. DE-AC05-76RL01830.

[1]  R. A. McKee, F. J. Walker, and M. F. Chisholm, Phys. Rev. Lett. **81**, 3014 (1998).
[2]  Z. H. Lim, N. F. Quackenbush, A. Penn, M. Chrysler, M. Bowden, Z. Zhu, J. M. Ablett, T. -L. Lee, J. M. LeBeau, J. C. Woicik, P. V. Sushko, S. A. Chambers and J. H. Ngai, Physical Review Letters **123**, 026805 (2019).
[3]  L. Kornblum, D. P. Fenning, J. Faucher, J. Hwang, A. Boni, M. G. Han, M. D. Morales-Acosta, Y. Zhu, E. I. Altman, M. L. Lee, C. H. Ahn, F. J. Walker and Y. Shao-Horn, Energy & Environmental Science **10**, 377 (2016).
[4]  L. Ji, M. D. McDaniel, S. Wang, A. B. Posadas, X. Li, H. Huang, J. C. Lee, A. A. Demkov, A. J. Bard, J. G. Ekerdt and E. T. Yu Nat. Nanotech. **10**, 84 (2015).
[5]  V. E. Henrich and P. A. Cox, *The Surface Science of Metal Oxides* (Cambridge University Press, Cambridge, 1994).
[6]  A. Ohtomo and H. Y. Hwang, Appl. Phys. Lett. **84**, 1716 (2004).
[7]  M. Chrysler, J. Gabel, T.-L. Lee, A. N. Penn, B. E. Matthews, D. M. Kepaptsoglou, Q. M. Ramasse, J. R. Paudel, R. K. Sah, J. D. Grassi, Z. Zhu, A. X. Gray, J. M. LeBeau, S. R. Spurgeon, S. A. Chambers, P. V. Sushko, and J. H. Ngai Physical Review Materials **5**, 104603 (2021).
[8]  S. A. Chambers and Y. Du, J. Vac. Sci. Technol. A **38**, 043409 (2020).
[9]  P. V. Sushko and S. A. Chambers, Sci. Rep. **10**, 13028, 13028 (2020).

[10] G. Kresse and J. Furthmüller, Phys. Rev. B **54**, 11169 (1996).
[11] G. Kresse and D. Joubert, Physical Review B **59**, 1758 (1999).
[12] J. P. Perdew, A. Ruzsinszky, G. I. Csonka, O. A. Vydrov, G. E. Scuseria, L. A. Constantin, X. L. Zhou, and K. Burke, Physical Review Letters **100**, 136406 (2008).
[13] P. E. Blöchl, Phys. Rev. B **50**, 17953 (1994).
[14] R. F. Bader, *Atoms in Molecules: A Quantum Theory* (Clarendon Press, 1990).
[15] W. Tang, E. Sanville, and G. Henkelman, J. Phys.: Condens. Matter **21**, 084204 (2009).
[16] We note that the slight downward band bending on the STO side of the buried interface is not physical, but rather an artefact. This artefact arises because of the high sensitivity of the fitting algorithm to slight differences between the fitting function and the heterojunction spectrum. In this case, the valley between the spin-orbit components in the heterojunction Sr 3d spectrum is slightly more filled in than that of the fitting function, which in turn comes from fitting a spectrum measured for bulk, flat-band $SrNb_{0.01}Ti_{0.99}O_3$(001) with a model spectrum consisting of pairs of Gaussians and Lorentzians. The algorithm responds by generating artificial band bending at the interface in order to minimize the cost function [see Ref. 9].
[17] A. Borghesi, B. Pivac, A. Sassella, and A. Stella, J. Appl. Phys. **77**, 4169 (1995).
[18] R. A. McKee, F. J. Walker, M. Buongiorno Nardelli, W. A. Shelton, and G. M. Stocks, Science **300**, 1726 (2003).
[19] See Supplemental Material at {link} for additional material characterization.
[20] V. E. Henrich, G. Dresselhaus, and H. J. Zeiger, Physical Review B **17**, 4908 (1978).
[21] V. E. Henrich, G. Dresselhaus, and H. J. Zeiger, J. Vac. Sci. Technol. **15**, 534 (1978).
[22] V. M. Bermudez and V. H. Ritz, Chem. Phys. Lett. **73**, 160 (1980).
[23] W. S. Samarakoon, P. V. Sushko, D. Lee, B. Jalan, H. Zhou, Y. Du, Z. Feng, and S. A. Chambers, Phys. Rev. Mater. **6**, 103405 (2022).
[24] Y. Aiura, I. Hase, H. Bando, T. Yasue, T. Saitoh, and D. S. Dessau, Surface Science **515**, 61, (2002).
[25] M. Andrä, Filip Dvořák, Mykhailo Vorokhta, Slavomír Nemšák, Vladimír Matolín, Claus M. Schneider, Regina Dittmann, and Felix Gunkel APL Materials **5**, 056106 (2017).
[26] J. Gabel, M. Pickem, P. Scheiderer, L. Dudy, B. Leikert, M. Fuchs, M. Stübinger, M. Schmitt, J. Küspert, G. Sangiovanni, J. M. Tomczak, K. Held, T.-L. Lee, R. Claessen, M. Sing Advanced Electronic Materials **8**, 2101006 (2022).
[27] X. Shen, K. Ahmadi-Majlan, J. H. Ngai, D. Wu, and D. Su, Appl. Phys. Lett. **106**, 032903 (2015).
[28] R. J. Cottier, N. A. Steinle, D. A. Currie, and N. Theodoropoulou, Applied Physics Letters **107**, 221601 (2015).
[29] D. Bao, X. Yao, N. Wakiya, K. Shinozaki, and N. Mizutani, Applied Physics Letters **79**, 3767 (2001).